\documentclass[graybox]{svmult}

\usepackage{type1cm}        
\usepackage{float}
\usepackage[caption=false]{subfig}
\usepackage{multirow}
\usepackage{amsmath,amssymb}
\usepackage{makeidx}         
\usepackage{graphicx}        
\usepackage{multicol}        
\usepackage[bottom]{footmisc}

\makeindex             

\begin{document}

\title*{Modified version of open TASEP with \\dynamic defects
}
\author{Nikhil Bhatia \& Arvind K. Gupta}
\institute{Nikhil Bhatia \at Department of Mathematics, Indian Institute of Technology Ropar,\\ Rupnagar, Punjab, India 140001 \email{nikhil.19maz0007@iitrpr.ac.in}
\and Arvind K. Gupta \at Department of Mathematics, Indian Institute of Technology Ropar,\\ Rupnagar, Punjab, India 140001 \email{akgupta@iitrpr.a.cin}}
%
%
\maketitle

\abstract*{Each chapter should be preceded by an abstract (no more than 200 words) that summarizes the content. The abstract will appear \textit{online} at \url{www.SpringerLink.com} and be available with unrestricted access. This allows unregistered users to read the abstract as a teaser for the complete chapter.
Please use the 'starred' version of the \texttt{abstract} command for typesetting the text of the online abstracts (cf. source file of this chapter template \texttt{abstract}) and include them with the source files of your manuscript. Use the plain \texttt{abstract} command if the abstract is also to appear in the printed version of the book.}

\abstract{We propose a modification to the study of site-wise dynamically disordered totally asymmetric simple exclusion process (TASEP). Motivated by the process of gene transcription, a study in ref. \cite{waclaw2019totally} introduced an extension of TASEP, where the defects (or obstacles) bind/un-bind dynamically to the sites of the lattice and the hopping of the particles on lattice faces a hindrance if the arrival site is occupied by an obstacle. In addition, the particle is only allowed to enter the lattice provided the first site is defect-free.  In our study, we propose that the particle movement at the entry of the lattice must face an equal hindrance that is provided by the obstacles to the rest of the particles on the lattice. For open boundaries, the continuum mean-field equations are derived and solved numerically to obtain steady-state phase diagrams and density profiles. The presence of obstacles produces a shift in the phase boundaries obtained but the same three phases as obtained for the standard TASEP. Contrary to the model introduced in ref. \cite{waclaw2019totally}, the idea to introduce the modification at the entrance shows that the limiting case $p_d \rightarrow 1$ converges to the standard TASEP, where $p_d$ refers to the affected hopping rate due to presence of obstacle. The mean-field solutions are validated using extensive Monte Carlo simulations.}

\section{Introduction}
\label{sec:1}
Consider a day-to-day traffic scenario where several incidents, such as vehicular crashes, debris, and traffic lights, are examples of physical obstructions that lead to vehicles bunching in the travel lane and disrupting the normal flow of traffic. Analogous situations also occur in the microscopic realm of molecular biology, where molecular motors moving through intracellular filaments, DNA/mRNA strands, or ions moving via ion channels are examples of ”vehicles.”

The stochastic transport in such situations is captured by the paradigmatic model totally asymmetric simple exclusion process (TASEP) \cite{krug1991boundary,katz1983phase}. In its simplest incarnation, TASEP was proposed to model biopolymerization, such as the synthesis of RNA on DNA templates \cite{macdonald1968kinetics,macdonald1969concerning}. From a theoretical standpoint, TASEP has been extensively studied as an archetype model of jamming, helped by the property that it is exactly solvable and that a mean-field approach gives the same result as the exact solution \cite{chou2004clustered,schadschneide$R_2$002traffic,klumpp2003traffic,de2005bethe,simon2009construction,chowdhury2005physics,derrida1998exactly,schutz1993phase,derrida1993exact}. Recently, models with dynamic defects have been studied, which, when present, slow down particles from moving down the chain of sites. Earlier, TASEP with defects was studied, where defects were present at only specific sites or sites with defects were randomly distributed \cite{janowsky1992finite,chaffey2003alberts,schliwa2003molecular,kolomeisky1998asymmetric,tripathy1997steady}. Several defect models are motivated by the biological process of ”roadblocking” during gene transcription or periodically switching traffic lights \cite{epshtein2003transcription,juhasz2006partially}. Recently, we investigated a TASEP model with site-wise dynamic defects, where the number of particles and the number of flaws in the system is both conserved \cite{bhatia2022finitedefect}. In this paper, we propose a modified version of TASEP with the dynamic disorder (ddTASEP), in which defects appear and disappear randomly and uniformly across the lattice. Both particles and defects join the lattice from an infinite reservoir. In contrast to the previously studied model \cite{waclaw2019totally}, the concept of an affected hopping rate of particles in the presence of defects has also been incorporated at the entry site, which has a significant effect on the stationary state properties of the system.

\section{Model description}
\label{sec:2}
\begin{figure}[t]
    \centering
    \includegraphics[width=5in,height=1in]{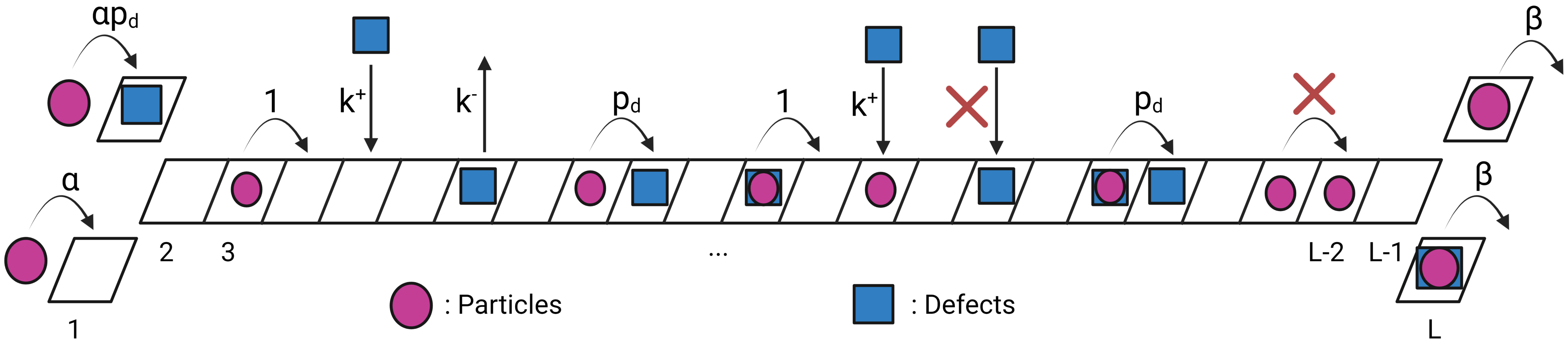}
    \caption{Schematic picture of the model (see section \ref{sec:2} for description).}
    \label{fig=model}
\end{figure}

We consider a modified version of the totally asymmetric simple exclusion process (TASEP) with site-wise dynamic defects (ddTASEP), where defects appear and disappear stochastically on any site $j = 1, \ldots, L$ of a one-dimensional lattice, as illustrated in Fig. \ref{fig=model}. Both defects and lattice individually follow the hard-core exclusion principle.
\begin{enumerate}
    \item {\bf Defect dynamics}: Particle occupancy has no effect on the dynamics of defects. Therefore defects behave uniformly throughout the lattice.
    \begin{enumerate}
        \item The defects can randomly bind a site without a defect with a rate $k^{+}$.
         \item The defects can randomly unbind from a site with a defect with a rate $k^{-}$.
    \end{enumerate}
    \item {\bf Particle dynamics}: The dynamics of particles are significantly influenced by defect occupancy. Hence these dynamics at various lattice locations are characterized as follows:
\begin{enumerate}
    \item {\bf At entry:} A particle enters the lattice through the first site ($j=1$) with a rate $\alpha$ if the defect is not present at the entry site. Else, it enters with a rate $\alpha p_d$, where $p_d<1$. 
    \item {\bf At bulk:} A particle hops from the site $j$ to $j + 1$ with a unit rate if the defect is unavailable at the target site $j + 1 $. Suppose the target site contains a defect; the particle hops with a rate $p_d$.
    \item {\bf At exit:} A particle present at the last site ($j=L$) leaves the lattice with a rate $\beta$.
\end{enumerate}
\end{enumerate}
Let $\sigma_j,\nu_j$ be discrete binary random variables such that $\sigma_j = 0 (1)$ denotes the absence (presence) of a particle at $j^{th}$ lattice site whereas $\nu_{j}=0 (1)$ denotes the absence (presence) of a defect at $j^{th}$ lattice site.
\section{Naive-mean field approximations for continuum equations}
The master equation for the temporal evolution of the average particle density in the bulk of the lattice ($2 \leq j \leq L - 1$) that substantiates analytical support to the process involved is given as,
\begin{equation}
    \frac{d \langle \sigma_{j} \rangle}{dt}=J_{j-1,j}- J_{j,j+1}, \label{Eq:1}
\end{equation}
where,
\begin{equation}
    J_{j-1,j}=\langle\sigma_{j-1}(1-\nu_j)(1-\sigma_{j})\rangle + p_d\langle\sigma_{j-1}\nu_j(1-\sigma_{j})\rangle
\end{equation}
is the particle flux from site $j$ to site $j+1$ and, $\langle\cdots\rangle$ stands for the statistical average. The following equation serves as the master equation for transitions at the left-lattice boundary:
\begin{equation}
    \frac{d\langle \sigma_{1} \rangle}{dt}= J_{entry}-J_{1,2},\label{Eq:2}
\end{equation}
where, $J_{entry}$ is the particle flow at the entry site and is provided as:
\begin{equation}
    J_{entry}=\alpha  \langle(1- \sigma_{1})(1-\nu_{1})\rangle + \alpha p_d  \langle(1- \sigma_{1})\nu_{1}\rangle, \label{Eq:entry}
\end{equation}
The second term in the equation above, which was not taken into account in the ref. \cite{waclaw2019totally}, is due to the consideration of defect binding at the left boundary.
The average particle occupancy number at the right-lattice border evolves in accordance with:
\begin{equation}
    \frac{d\langle \sigma_{L} \rangle}{dt}= J_{L-1,L}-J_{exit},\label{Eq:3}
\end{equation}
where $J_{exit}$ denotes the particle current at the exit site ($j=L$), and it is given as:
\begin{equation}
    J_{exit}=\beta \langle \sigma_{L} \rangle. \label{Eq:exit}
\end{equation}
Additionally, the following equation provides the evolution of the average defect occupancy number in the lattice:
\begin{equation}
        \frac{d \langle \nu_{j} \rangle}{dt}=k^{+}_{eff}\langle (1-\nu_{j})\rangle - k^{-}\langle \nu_{j}\rangle,  \,\,\,\,\, 1\leq j\leq L \label{Eq:4}.  
\end{equation}
We now use naive mean-field approximation, which ignores one and two-point correlators in the generated master equations, i.e., $\langle \sigma_j \sigma_{j+1}\rangle = \langle \sigma_j \rangle \langle \sigma_{j+1}\rangle $. We also assume that the binding/un-binding rates are of the order $k^{+},k^{-} \gtrsim 1$, implying that $\langle \sigma_j\nu_{j+1}\rangle = \langle \sigma_j\nu_{j+1}\rangle$. For the continuum limit of the model, we coarse-grain the lattice by introducing a quasi-continuous variable $x=\epsilon_j \in [0, 1]$ and replace the discrete binary variables $\sigma_j, \nu_j$ with continuous variables $\rho_j, \rho_{d, j} \in [0, 1]$, where $\epsilon=\frac{1}{L}$ is a lattice constant and $t'=\frac{t}{L}$ is a re-scaled time. The consideration of spatial homogeneity on the lattice along with the application of the Taylor series of $\rho (x \pm \epsilon)$ in Eq. \eqref{Eq:1} reforms Eq. \eqref{Eq:4} and Eq. \eqref{Eq:1} into the system of differential equations governing the state of the system:
\begin{subequations}
    \begin{align}
\frac{\partial \rho}{\partial t'}  &=\frac{\partial J}{\partial x}\label{Eq:5a}\\
\frac{\partial \rho_d}{\partial t'} &= k^{+} (1-\rho_{d}) - k^{-} \rho_{d} \label{Eq:5b}
\end{align}
\end{subequations}
For finite $\epsilon$, $J=(1-\rho_d+p_d\rho_d)\big(\frac{\epsilon}{2} \frac{\partial \rho}{\partial x} + \rho (1-\rho)\big)$ denotes the average current whereas in the continuum limit $\epsilon \rightarrow 0^+$, it becomes $J=(1-\rho_d+p_d\rho_d)\rho(1-\rho)$.
\begin{figure*}[t]
\centering
\subfloat[$(\alpha=0.3,\beta=0.7)$]{\includegraphics[width=1.6in]{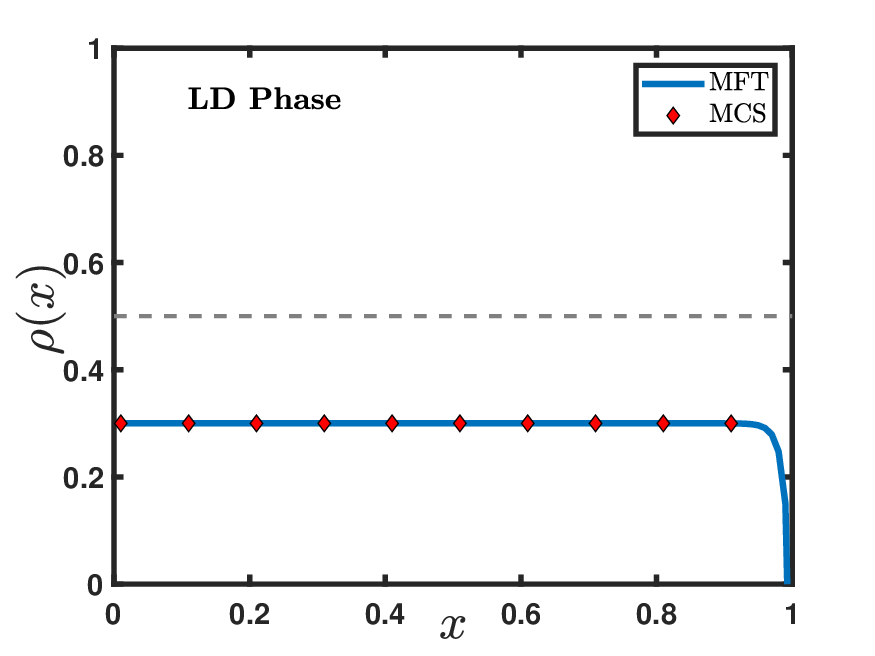}}
\subfloat[$(\alpha=0.7,\beta=0.7)$]{\includegraphics[width=1.6in]{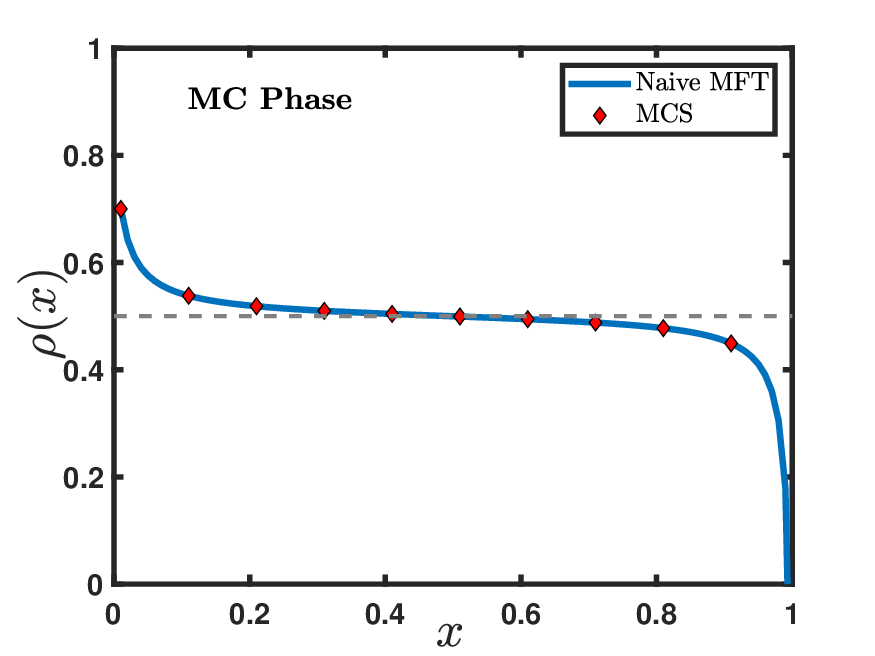}}
\subfloat[$(\alpha=0.7,\beta=0.2)$]{\includegraphics[width=1.6in]{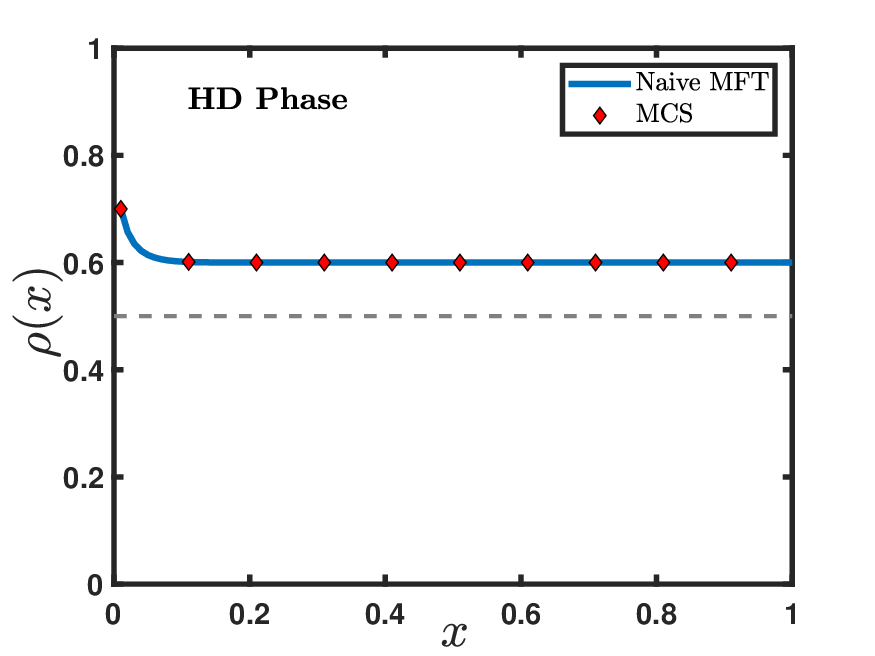}}
\caption{Density profiles for the system with open boundaries and random sequential update rule of a (a) LD, (b) HD, (c) MC phase, where $z=0.5$ corresponding to $k^{+}=k^{-}=5$ and $p_d=0$. All figures use solid lines to show findings from the naïve mean-field approximation, and symbols to show results from Monte Carlo simulations with $L = 100$. To show the particle density in comparison to the density in the MC phase, a grey dotted line with a value of $1/2$ is drawn.}
\label{fig:fig2}
\end{figure*}
\section{Steady-state analysis}
At stationary state, the non-linear differential equation \eqref{Eq:5a} in the limit $\epsilon=0$ (valid for large system size $L$) yields a first order differential equation,
\begin{equation}
   (1-\rho_d+p_d\rho_d)(1-2\rho) \frac{\partial \rho}{\partial x}=0 \label{Eq:6},
\end{equation}
whereas the steady-state defect density from Eq. \eqref{Eq:5b} is computed as,
\begin{equation}
    \rho_d=\frac{k^{+}}{k^{+} +k^{-}}. \label{Eq:7}
\end{equation}
For the sake of simplicity and to reduce the parameter space, let us define $z=\rho_d(1-p_d)$ as the obstruction factor that quantifies the particle's hindrance on the lattice.
Similarly, the continuity equation for boundary currents via a mean-field approximation determines the boundary density at left-boundary and right-boundary density as:
\begin{equation}
    \rho_{1}=\alpha,  \label{Eq:8}
\end{equation}
and,
\begin{equation}
    \rho_{L}=1-\frac{\beta}{1-z}. \label{Eq:9}
\end{equation}
respectively.
\begin{figure*}[t]
\centering
\subfloat[]{\includegraphics[width=2.4in]{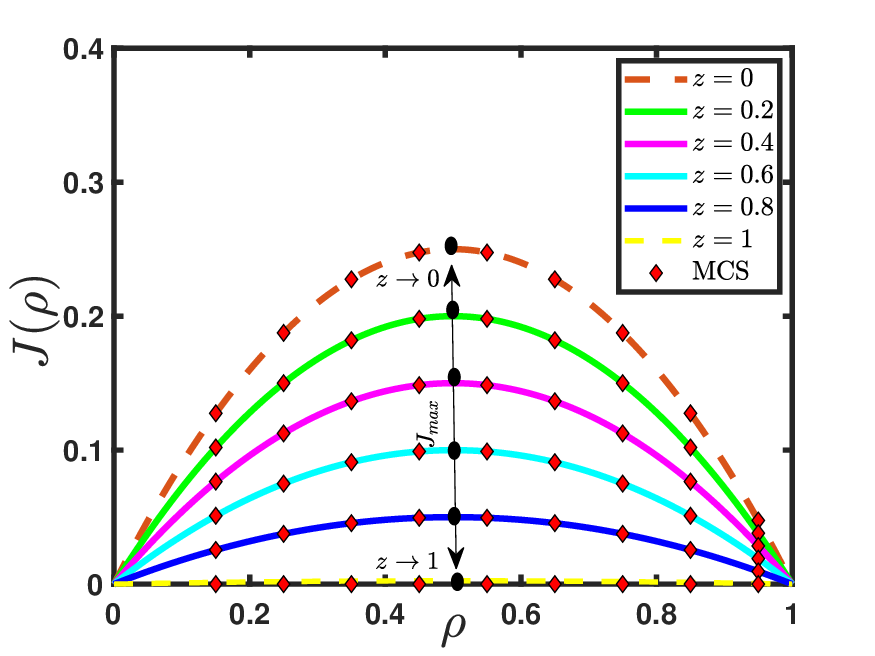}}
\subfloat[]{\includegraphics[width=2.4in]{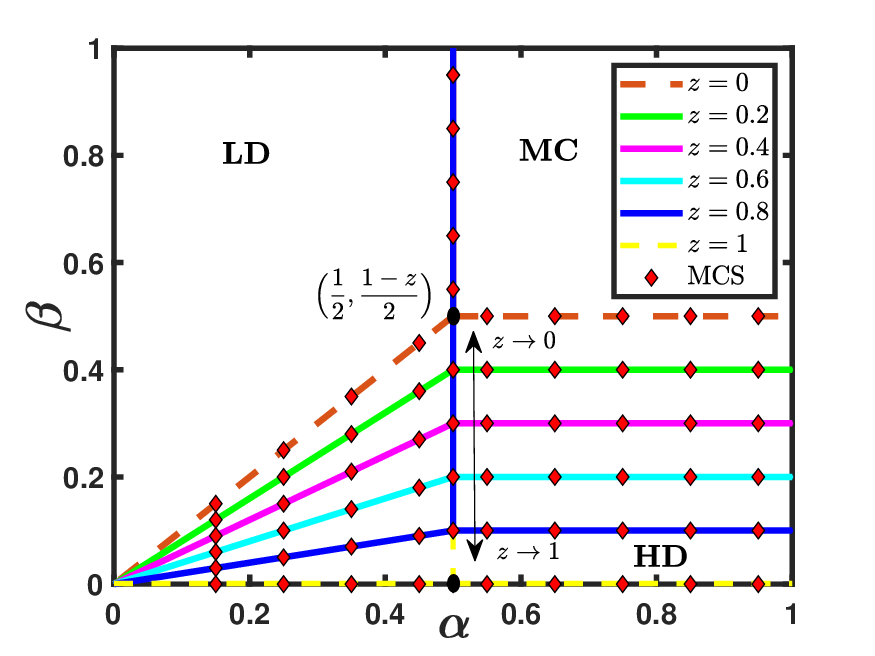}}
\caption{The effect of $z$ on (a) the bulk as well as the maximal current on lattice, (b) the phase plane obtained through naive mean-field approach for $k^{+},k^{-} \gtrsim 1$.}
\label{fig:fig3}
\end{figure*}
The phase boundaries computed utilizing the extremal current principle are as follows, along with the particle density in the associated phase region:
\begin{equation}
    \rho(x)=\begin{cases}
\alpha;& \alpha < \min\big(\frac{\beta}{1-z}, \frac{1}{2}\big),\\ 
1-\frac{\beta}{1-z};& \frac{\beta}{1-z}< \min\big(\alpha, \frac{1}{2}\big),\\
\frac{1}{2};& \min\big(\alpha, \frac{\beta}{1-z} \big) \geq \frac{1}{2}. \label{Eq:10}\\
\end{cases}    
\end{equation}
The density profiles for a fixed choice of $\alpha, \beta$ and $z$ are given in Fig. \ref{fig:fig2}. We now examine the system's stationary characteristics and try to understand the effect of the obstruction factor on the bulk current, followed by the $\alpha-\beta$ parameter space. The current-density relation $J(\rho)=(1-z) \rho (1-\rho)$, as well as Fig. \ref{fig:fig3} (a), clearly show that the bulk current for the TASEP with the dynamic disorder in the proposed model is an inverted parabola, with the maximal current given by $J_{max}=J(\frac{1}{2})=\frac{1-z}{4}$ being a decreasing function of $z$. The Fig. \ref{fig:fig3} (b) depicts the effect of $z$ on the phase plane and illustrates that as $z$ decreases, the particle encounters less resistance, and the triple point coordinates $(\frac{1}{2}, \frac{1-z}{2})$ move vertically upwards, raising the slope of the phase boundary $\alpha=\frac{\beta}{1-z}$ and the position of the phase boundary $\beta = \frac{1-z}{2}$. As a result, the LD and MC phases contract while the HD phase expands. In the case of $z \rightarrow 0$, the impact of obstruction on the lattice completely disappears, causing the triple point and phase plane to converge to that of a standard TASEP with infinite resources.
However, when $z$ increases, the particle encounters more obstruction, and the triple point moves vertically downwards, causing the slope of the phase boundary $\alpha=\frac{\beta}{1-z}$ to fall and the position of the phase boundary $\beta = \frac{1-z}{2}$ moves vertically downward. As a result, the HD phase reduces while the LD and MC phases grow. The obstruction factor achieves its greatest value in the limit $z \rightarrow 1$, resulting in a zero steady-state current in the system. Furthermore, the triple point's coordinates for the limiting situation, $z \rightarrow 1$, approach the $\alpha$-axis, causing the HD phase to vanish completely. The LD phase and MC phase will be the only two phases present in the phase plane, as shown in Fig. \ref{fig:fig3} (b).

\section{Conclusion}
We examined a single-channel TASEP model with a dynamic disorder where the particle dynamics at the entry site are modified to incorporate the effect of defects throughout the lattice uniformly. The steady-state properties of the system are studied using a mean-field approximation in the continuum limit. We investigated the role of site-wise dynamic defects utilizing an obstruction factor that unifies the functions of parameters $p_d$ and $\rho_d$. It allows us to collectively study their effect on the system's stationary properties. We have obtained the explicit expression of boundary densities and phase boundaries. The limiting cases $z \rightarrow 0 $ and $z \rightarrow 1$ are dealt with to understand better how obstructions affect the system's stationary state.
In contrast to the ref. \cite{waclaw2019totally}, the modification of the particle dynamics at the entry site due to consideration of the defect's presence at the first site turns out to be more realistic and constructive because the topology of the phase diagram now converges to the standard open-TASEP with infinite resources in the limit $z \rightarrow 0$ (or $p_d \rightarrow 1$). On varying the obstruction factor, the only changes in the structure of the phase diagram found are the gradual shifting of the phase boundaries and the shrinkage/expansion of various phases. The Monte Carlo simulation is used to validate the outcomes of the continuum mean-field equations. Our theoretical research aimed to simulate dynamic aspects of potential defect-prone transport mechanisms and shed light on their stationary qualities. The research may aid in understanding the intricate dynamics of numerous physical and biological systems. 
\label{sec:4}
\begin{acknowledgement}
The first author thanks the Council of Scientific and Industrial Research (CSIR), India, for financial support under File No:09/1005(0028)/2019-EMR-I, and AKG acknowledges support from DST-SERB, Govt. of India (Grant CRG/2019/004669 \& MTR/2019/000312). This work was partially supported by the FIST program of the Department of Science and Technology, Government of India, Reference No. SR/FST/MS-I/2018/22(C).
\end{acknowledgement}
\addcontentsline{toc}{section}{Appendix}
%
%

\bibliographystyle{elsarticle-num}

\end{document}